\documentclass[
 twocolumn,
 amsmath,
 amssymb,
 aps,
 prb,
]{revtex4-2} 

\usepackage{graphicx}
\usepackage{dcolumn}
\usepackage{multirow}
\usepackage{adjustbox}
\usepackage{amsmath}
\usepackage{bm}

\usepackage[utf8x]{inputenc}
\usepackage{pslatex}
\usepackage{textcomp} 
\usepackage{hyperref}
\usepackage{lmodern} 
\usepackage{breakurl} 

\begin{document}
\begin{sloppypar} 

\title{Structural transformation and magnetic properties of (Fe$_{0.7}$Co$_{0.3}$)$_2$B alloys doped with 5\textit{d} elements: A combined first-principles and experimental study}

\author{A. Musiał}
 \email[Corresponding author: ]{musial@ifmpan.poznan.pl}
 \address{Institute of Molecular Physics, Polish Academy of Sciences, M. Smoluchowskiego 17, 60-179 Poznań, Poland, Center for Advanced Technologies, Adam Mickiewicz University, Uniwersytetu Poznańskiego 10, 61-614 Poznań, Poland}
 
\author{W. Marciniak}

\address{Institute of Physics, Faculty of Technical Physics, Poznań University of Technology, Piotrowo 3, 61-138 Pozna\'{n}, Poland, \\ Institute of Molecular Physics, Polish Academy of Sciences, M. Smoluchowskiego 17, 60-179 Poznań, Poland}

\author{Z. Śniadecki}
\author{M. Werwiński}
\author{P. Kuświk}
\author{B. Idzikowski}

\address{Institute of Molecular Physics, Polish Academy of Sciences, M. Smoluchowskiego 17, 60-179 Poznań, Poland}

\author{M. Kołodziej}

\address{Institute of Molecular Physics, Polish Academy of Sciences, M. Smoluchowskiego 17, 60-179 Poznań, Poland, \\ NanoBioMedical Centre, Adam Mickiewicz University in Poznań, Wszechnicy Piastowskiej 3, 61-614 Poznań, Poland}

\author{A. Grabias}
\author{M. Kopcewicz}

\address{Łukasiewicz – Institute of Microelectronics and Photonics, Lotników 32/46, 02-668 Warsaw, 
Center of Electronic Materials Technology, Wólczyńska 133, 01-919 Warsaw, Poland}

\author{J. Marcin}
\author{J. Kováč}

\address{Institute of Experimental Physics, Slovak Academy of Sciences, 
Watsonova 47, 040 01 Košice, Slovakia}

\begin{abstract}

(Fe,Co)$_2$B-based compounds with specific 5\textit{d} substitutions are considered as promising materials for permanent magnets without rare-earth elements.
We conducted a combined first-principles and experimental study focused on (Fe$_{0.7}$Co$_{0.3}$)$_2$B alloys doped with W and Re.
First, we used full-potential local-orbital scheme to systematically investigate (Fe,Co)$_2$B alloys with 3$d$, 4$d$, and 5$d$ substitutions.
Computational analyses showed a significant increase in magnetocrystalline anisotropy only for the Re doped sample. 
Simultaneously, the structural and magnetic properties of the (Fe$_{0.7-x}$Co$_{0.3-x}$M$_{2x}$)$_2$B ($M$ = W, Re; $x$ = 0, 0.025) alloys were  investigated experimentally.
The desired (Fe,Co)$_2$B tetragonal phase was synthesized by heat treatment of amorphous precursors. 
We observed that isothermal annealing increases the coercive field of all samples.
However, the obtained values, without further optimization, are well below the threshold for permanent magnet applications.
Nevertheless, annealing of substituted samples at 750$^o$C significantly improves saturation magnetization values.
Furthermore, M\"{o}ssbauer spectroscopy revealed a reduction of the hyperfine field due to the presence of Co atoms in the (Fe,Co)$_2$B phase, where additional defect positions are formed by Re and W. 
Radio-frequency M\"{o}ssbauer studies showed that (Fe$_{0.7}$Co$_{0.3}$)$_2$B and the W-substituted sample began to crystallize when exposed to a radio frequency field of 12~Oe, indicating that the amorphous phase is stabilized by Re substitution.
Improvement of thermal stability of (Fe$_{0.675}$Co$_{0.275}$Re$_{0.05}$)$_2$B alloy is consistent with the results of differential scanning calorimetry and thermomagnetic measurements.\\

\begin{description}
\item[Keywords] 
amorphous alloy, magnetic properties, magnetocrystalline anisotropy energy, crystal structure, radio-frequency M\"{o}ssbauer spectroscopy.

\end{description}
\end{abstract}

\maketitle

\section{Introduction}

Permanent magnets are one of the most important materials used in modern technology.
The best known and widely used permanent magnets are characterized by high $T_C$ values, such as SmCo$_{5}$ alloys~\cite{das_anisotropy_2019}, and high energy product (BH)$_{\mathrm{max}}$,  where the most obvious example is Nd$_{2}$Fe$_{14}$B \cite{toga_anisotropy_2018}.
As rare earth prices have shown high volatility over the past decade, the search for high-energy product magnets with limited content of elements such as Sm or Nd has intensified~\cite{cui_current_2018, skokov_heavy_2018, mohapatra_chapter_2018}.
Currently, candidates for rare-earth-free permanent magnets include for example:
FeNi compounds~\cite{sakurai_metastable_2020},
Fe$_{16-x}$Co$_x$N$_2$ alloys~\cite{zhao_large_2016},
and manganese-based materials such as MnBi, MnAl, and MnGa~\cite{patel_rare-earth-free_2018, Jia2020}.
Promising candidates, recently predicted from data-mining approach, are also Pt$_2$FeNi, Pt$_2$FeCu, and W$_2$FeB$_2$~\cite{vishina_high-throughput_2020}.
%
%
The prospect of use as rare-earth-free permanent magnets has also sparked interest in (Fe$_{1-x}$Co$_{x}$)$_2$B alloys.
Experimental studies of these alloys include, for example,
analyses of the full range of cobalt concentration~\cite{edstrom_magnetic_2015,belashchenko_origin_2015,wallisch_synthesis_2015} 
along with reinvestigation of their intrinsic magnetic properties~\cite{lamichhane_reinvestigation_2020}.
Other experiments involved characterization of (Fe$_{0.7}$Co$_{0.3}$)$_2$B single crystal~\cite{kuzmin_towards_2014} and
mechanically milled (Fe$_{0.675}$Co$_{0.3}$Re$_{0.025}$)$_2$B~\cite{kim_coercivity_2018}.
Experimental efforts were also supported by numerous first-principles calculations.
For example, the calculations have helped to explain the origin of spin-reorientation transitions in (Fe$_{1-x}$Co$_{x}$)$_2$B alloys~\cite{belashchenko_origin_2015} and
spin-fluctuation mechanism of the temperature dependence of the magnetocrystalline anisotropy of (Fe$_{1-x}$Co$_{x}$)$_2$B~\cite{zhuravlev_spin-fluctuation_2015}.
The calculations also predict a high-pressure induced magnetic moment collapse~\cite{gueddouh_magnetic_2018}.
Moreover, some groups have investigated the effect of doping with elements 3$d$~\cite{wei_effect_2018} and 5$d$~\cite{edstrom_magnetic_2015}.
Another computational work involved accurate estimation of the magnetocrystalline anisotropy energies for (Fe$_{1-x}$Co$_{x}$)$_2$B alloys~\cite{dane_density_2015}.
In several cases, the results on (Fe$_{1-x}$Co$_{x}$)$_2$B alloys were the effect of synergy between experiment and theory~\cite{kuzmin_towards_2014, edstrom_magnetic_2015, belashchenko_origin_2015}. 
Furthermore, also a number of related boride phases, such as
Co$_{2-x}$Mn$_x$B, 
(Fe$_x$Co$_{1-x}$)$_3$B, and
Mn$_{0.95-\delta}$Fe$_{1.05-\delta-x}$Co$_x$B~\cite{pal_properties_2017,ener_magnetic_2019,abramchuk_tuning_2019},
have been studied recently
to tune their magnetic properties.
%
In this work, we investigate the rare-earth-free magnet (Fe$_{0.7}$Co$_{0.3}$)$_2$B and its alloys substituted with Re and W.
The Co content and the type of dopant elements were predicted in a previous study~\cite{edstrom_magnetic_2015} in which one of us participated.
The purpose of this work is to verify the previous predictions using a more advanced computational method based on a full-potential approach and to experimentally characterize the structural and magnetic properties of the alloys. 
Density functional theory along with X-ray diffraction, differential scanning calorimetry, macroscopic magnetic measurements, and M\"{o}ssbauer spectroscopy were utilized to reach this aim.
%
Based on the experimental work mentioned above, we will present the basic physical properties of Fe$_2$B, Co$_2$B, (Fe$_{0.7}$Co$_{0.3}$)$_2$B, and W- and Re-substituted alloys.
Fe$_2$B and Co$_2$B crystallize in a tetragonal structure with space group $I$4/$mcm$~\cite{edstrom_magnetic_2015}.
The experimental value of the magnetic anisotropy constant \textit{K}$_1$ is equal to -0.80~MJ\,m$^{-3}$ for Fe$_2$B and 0.10~MJ\,m$^{-3}$ for Co$_2$B~\cite{iga_magnetocrystalline_1970}.
The total magnetic moment drops from about 1.9~$\mu_B$/3$d$~atom for Fe$_2$B to 0.8~$\mu_B$/3$d$~atom for Co$_2$B~\cite{cadeville_nuclear_1975}.
The Curie temperature for Fe$_2$B is above 1000~K \cite{edstrom_magnetic_2015}, while that for Co$_2$B is equal to 433~K~\cite{cadeville_nuclear_1975}.
%
%
The (Fe$_{0.7}$Co$_{0.3}$)$_2$B alloy shows the highest value of magnetocrystalline anisotropy constant among the alloys containing tetragonal (Fe,Co)$_2$B phase~\cite{iga_magnetocrystalline_1970,edstrom_magnetic_2015} and one of the highest values of magnetic moment.
Its Curie temperature is relatively high and equal to above 800~K~\cite{edstrom_magnetic_2015}.
The addition of 2.5~at.\% Re in place of the 3$d$ elements results in an increase in the magnetocrystalline anisotropy constant of the alloy (Fe$_{0.7}$Co$_{0.3}$)$_2$B from 0.42 to 0.63~MJ\,m$^{-3}$ (at room temperature) and an increase of anisotropy field from 1~T to 1.6~T.
At the same time, the saturation magnetization (at 2~T) decreases from 143.3 to 122.2~A\,m$^2$\,kg$^{-1}$~\cite{edstrom_magnetic_2015}.

\section{Computational and Experimental Details}\label{sec:exp_comp_details}
\subsection{Computational Details}\label{subsec:comp_details}
\begin{figure}[t!]
\centering
\includegraphics[clip, width = 0.8 \columnwidth]{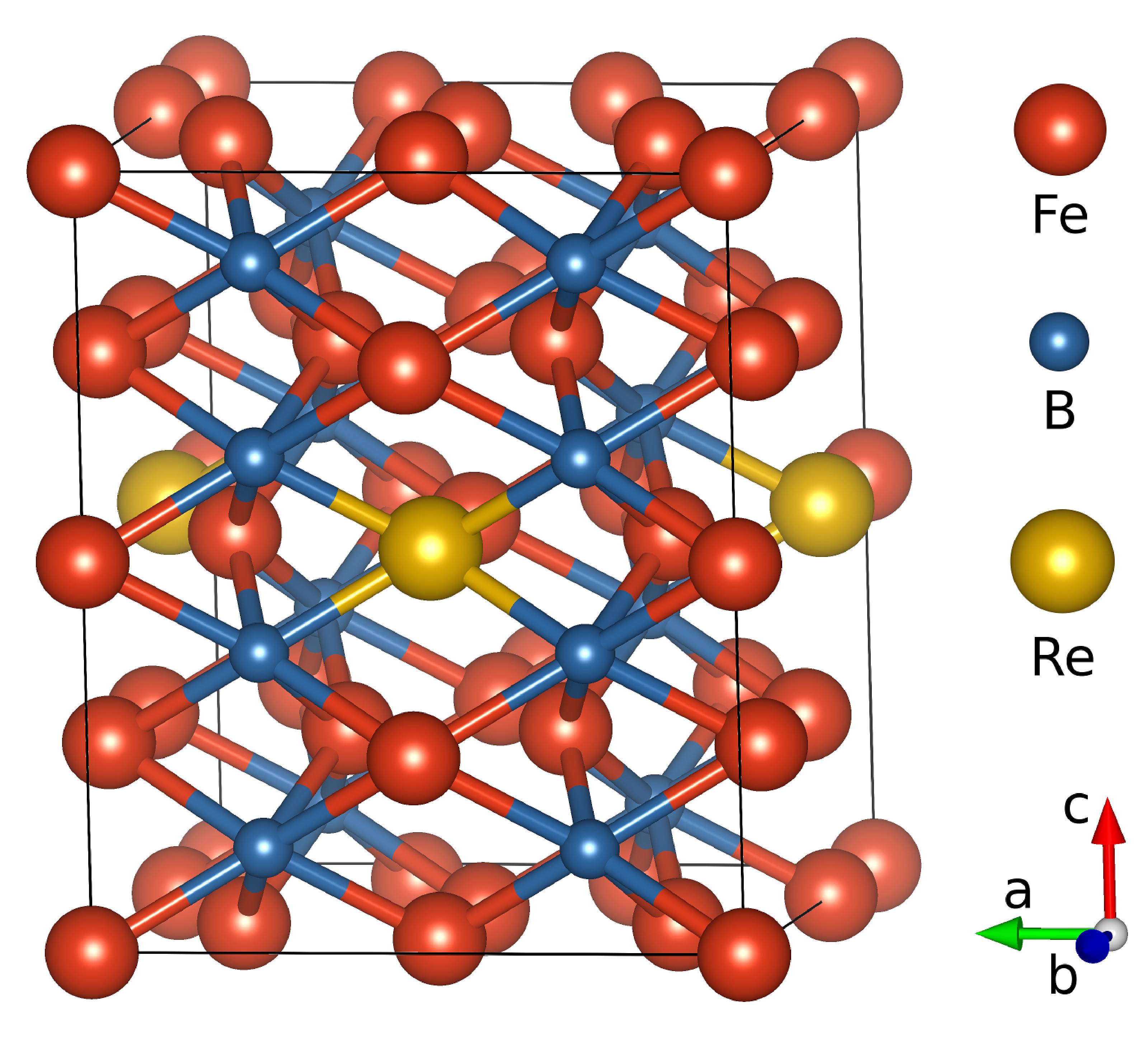}
\caption{
Supercell of Fe$_{15}$ReB$_8$ (based on tetragonal Fe$_2$B structure) with 
space group $I$4/$mcm$ ($a$~=~7.116~\AA{}, $c$~=~8.482~\AA{}).
The presented model consists of sixteen formulas of Fe$_2$B.
}
\label{fig_struct} 
\end{figure}

%
%
Calculations of the electronic band structure were carried out using the 
full-potential local-orbital electronic structure code FPLO18.00-52~\cite{koepernik_full-potential_1999} employing a fixed atomic-like basis set.
FPLO was the choice for accurate magnetocrystalline anisotropy energy (MAE) calculations due to the full potential and fully-relativistic implementation of the code.
For the exchange-correlation potential, we used the generalized gradient approximation (GGA) in the Perdew-Burke-Ernzerhof (PBE) form~\cite{perdew_generalized_1996}.
For the scalar-relativistic calculations, we used the criterion of the convergence of charge density equal to 10$^{-6}$.
%
%
MAE was evaluated as the difference between fully-relativistic total energies calculated in one iteration for the quantization axes [001] and [100].
We confirmed the reliability of such a procedure by performing the full self-consistent calculations for selected cases.
The positive MAE sign corresponds to the axis of easy magnetization along the direction [001] ($c$ axis), see Fig.~\ref{fig_struct}.
%
%
For a selected alloy doped with Re, we examined the MAE as a function of spin magnetic moment ($m_s$) 
 using the fully-relativistic version of the fixed spin moment (FSM) approach~\cite{schwarz_itinerant_1984, werwinski_magnetic_2016}.
%
%
For the supercell models under consideration, the energy convergence with the number of $\mathbf{k}$-points was thoroughly tested.
We found that a $12 \times 12 \times 12$ \textbf{k}-mesh leads to the well converged MAE results.

%
The exemplary model of (Fe$_{0.66}$Co$_{0.28}$Re$_{0.06}$)$_2$B is obtained by combining 
a supercell approach with virtual crystal approximation (VCA).
%
First, based on the tetragonal Fe$_2$B unit cell, we prepared the Fe$_{15}$ReB$_8$ supercell consisting of eight Fe$_2$B formulas and with a single Fe atom replaced by Re, see~Fig. \ref{fig_struct}.
This procedure leads to a crystalline structure containing 11 nonequivalent atomic positions (8~Fe, 2~B, and 1~Re).
Next, in place of the remaining Fe atoms, we introduced the virtual atoms with a fractional atomic number 26.3, in order to mimic the 
70\% concentration of Fe (with atomic number Z~=~26) and 
30\% concentration of Co (with atomic number Z~=~27).
Because one Fe atom from the supercell was previously replaced by Re,
the final composition can be written as (Fe$_{0.66}$Co$_{0.28}$Re$_{0.06}$)$_2$B.
Subsequently, by replacing the Re atoms with other transition metals, we prepared the models containing 3$d$, 4$d$, and 5$d$ elements.
The crystallographic parameters for (Fe$_{0.7}$Co$_{0.3}$)$_{15}$MB$_8$ supercells ($a$~=~7.116~\AA{}, $c$~=~8.482~\AA{}) were based on the parameters for (Fe$_{0.7}$Co$_{0.3}$)$_2$B ($a$~=~5.032~\AA{}, $c$~=~4.241~\AA{}), which in turn  were interpolated based on the theoretically predicted parameters of Fe$_2$B ($a$~=~5.059~\AA{}, $c$~=~4.239~\AA{}) and Co$_2$B ($a$~=~4.969~\AA{}, $c$~=~4.244~\AA{}).
%
%
The VESTA code was used for visualization of the crystal structure~\cite{momma_vesta_2008} presented in Fig. \ref{fig_struct}.


\subsection{Experimental details}\label{subsec:comp_details}

Master alloys of (Fe$_{0.7-x}$Co$_{0.3-x}$M$_{2x}$)$_2$B ($M$ = W, Re; $x$ = 0, 0.025) were prepared by arc-melting technique. High purity elements Fe, Co, and B (3N or more) were re-melted several times in the argon atmosphere to obtain full homogeneity. The mass of ingots was controlled at each step of the synthesis in order to maintain the nominal composition. The alloys were subsequently injected on the copper wheel rotating with the surface velocity of 30 m\,s$^{-1}$ in an argon atmosphere. The  melt-spun ribbons were 30 $\mu$m thick. Densities of the alloys were calculated assuming the density of Co$_{53}$B$_{27}$ equal to 8.08 g$\,$cm$^{-1}$ \cite{Hasegawa1979}. Calculated densities are 7.53, 7.97, and 8.04 g$\,$cm$^{-1}$ for (Fe$_{0.7}$Co$_{0.3}$)$_2$B, (Fe$_{0.675}$Co$_{0.275}$W$_{0.05}$)$_2$B, and (Fe$_{0.675}$Co$_{0.275}$Re$_{0.05}$)$_2$B, respectively. Structural information was obtained by X-ray diffraction (XRD) with the use of TUR-M62 diffractometer (HZG4 goniometer) with CoK$_\alpha$ radiation ($\lambda$ = 1.7889 Å) in Bragg-Brentano geometry. The thermal stability of as-quenched samples was investigated by differential scanning calorimetry (DSC), in the temperature range from 100 to 850$^o$C, at a heating rate of 10 K min$^{-1}$, using Netzsch DSC 404 apparatus. Magnetic hysteresis loops were measured using a vibrating sample magnetometer (VSM) option in the Quantum Design Physical Property Measurement System (PPMS). Magnetization measurements at high temperatures were collected using a homemade VSM. In order to obtain information on the local $^{57}$Fe environment, $^{57}$Fe M\"{o}ssbauer spectroscopy measurements were made at room temperature. The hyperfine structure was modeled by least-square fitting using NORMOS program \cite{Brand1983}.
An unconventional rafio-frequency-M\"{o}ssbauer technique was used to study the effect of the chemical composition on the physical properties of amorphous alloys. The spectra were measured when the samples were exposed to a radio-frequency magnetic field from 0 to 18 Oe at 61.8 MHz.

\section{Results and Discussion}\label{sec:results}


\subsection{Density functional theory calculations}

%
\begin{figure}[!t]
\centering
\includegraphics[clip, width=0.98\columnwidth]{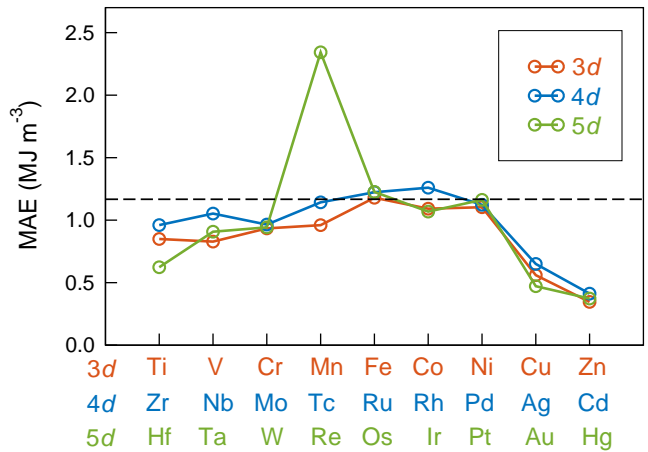}
\caption{
Magnetocrystalline anisotropy energy (MAE) for (Fe$_{0.66}$Co$_{0.28}$M$_{0.06}$)$_2$B alloys, where M stands for 3$d$, 4$d$, and 5$d$ elements.
The calculation was performed using FPLO18 code with PBE functional.
The dashed line means the calculated MAE of (Fe$_{0.7}$Co$_{0.3}$)$_2$B (1.17~MJ\,m$^{-3}$). 
The alloy model is prepared using a supercell combined with a virtual crystal approximation (VCA).
}
\label{fig:fecoxb_3d_4d_5d_MAE}
\end{figure}

(Fe$_{0.7}$Co$_{0.3}$)$_2$B shows the highest value of magnetocrystalline anisotropy among the (Fe,Co)$_2$B tetragonal alloys~\cite{iga_magnetocrystalline_1970,edstrom_magnetic_2015}.
To further increase the MAE of this material, one strategy is to combine it with elements from the $d$-block, particularly from 5$d$-series~\cite{edstrom_magnetic_2015}.
Previously, when examining the impact of the 5$d$ substitutions from the first principles, by applying atomic sphere approximation (ASA), Edstr\"om and coworkers \cite{edstrom_magnetic_2015} stated that the positive effect on the MAE of (Fe$_{0.7}$Co$_{0.3}$)$_2$B should be due to the doping of W and Re.
While the growth of MAE due to the alloying with Re was confirmed experimentally, problems with the preparation of single-phase samples with W prevented from evaluating this prediction~\cite{edstrom_magnetic_2015}.
In this work, we perform similar calculations, but using the full-potential approach instead atomic sphere approximation and extending the range of
dopants tested. Moreover, we were successful in synthesis of single phase W-containing alloy, what enabled us to make comprehensive analysis.
Figure~\ref{fig:fecoxb_3d_4d_5d_MAE} shows the results of our calculations of the influence of $3d$-, $4d$-, and $5d$-substitutions on the MAE of (Fe$_{0.7}$Co$_{0.3}$)$_2$B.
We see a maximum of MAE in the middle of each series, but a significant increase in MAE was observed only for (Fe$_{0.66}$Co$_{0.28}$Re$_{0.06}$)$_2$B.
However, given the above arguments, we decided that both compositions, with W and with Re, would be subjected to thorough theoretical and experimental analysis.

\begin{figure}[!t]
\centering
\includegraphics[clip, width=0.95\columnwidth]{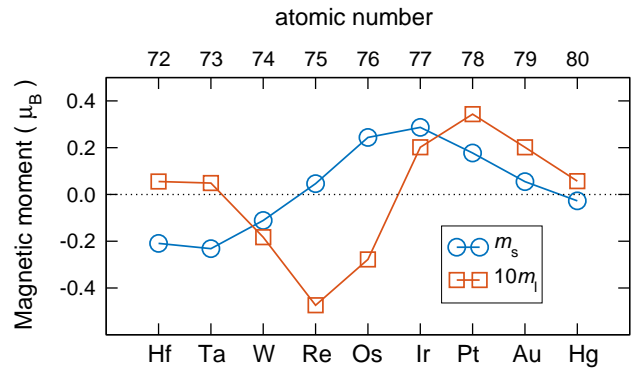}
\caption{
Spin ($m_\mathrm{s}$) and orbital ($m_\mathrm{l}$) magnetic moments of 5$d$ transition metal impurities M in (Fe$_{0.66}$Co$_{0.28}$M$_{0.06}$)$_2$B alloys 
as calculated for spin quantization axis along $c$-axis.
The calculation was performed using FPLO18 code with PBE functional.
}
\label{fig:fecoxb_5d_mm}
\end{figure}

It has been shown previously, that the growth in the MAE seen for W and Re dopants is primarily due to the strong spin-orbit coupling of 5$d$ element. However, other changes in the electronic structure also have an impact on MAE~\cite{edstrom_magnetic_2015,werwinski_magnetocrystalline_2018}.
The calculations from this work predict that the MAE increases from 1.17~MJ\,m$^{-3}$ for (Fe$_{0.7}$Co$_{0.3}$)$_2$B to about 2.34~MJ\,m$^{-3}$ for (Fe$_{0.66}$Co$_{0.28}$Re$_{0.06}$)$_2$B.
Although the qualitative results obtained are reliable, the exact values we have determined are subject to a known error of the virtual crystal approximation (VCA).
Thus, we estimate that the exact MAE values are overestimated by about 2 - 4 times compared to experiment~\cite{edstrom_magnetic_2015}.
Before we proceed to a detailed analysis of the compositions with W and Re, we would like to present how the magnetic moments induced on 5$d$ dopants change along the 5$d$-series.
Figure~\ref{fig:fecoxb_5d_mm} shows the calculated spin and orbital magnetic moments on 5$d$ transition metal impurities M in (Fe$_{0.66}$Co$_{0.28}$M$_{0.06}$)$_2$B alloys.
The spin magnetic moments of the early 5$d$ impurities (Hf, Ta, W) are coupled antiferromagnetically to the Fe/Co moments, while the moments of the late 5$d$ elements are coupled to Fe/Co ferromagnetically, except Hg.
The calculated spin moments take values between approximately -0.2 and 0.3~$\mu_{\mathrm{B}}$\,atom$^{-1}$.
The above results can be compared with the previously studied behavior of the 5$d$ impurities in the Fe matrix~\cite{wienke_determination_1991}.
The measured local magnetic moments of 5$d$ impurities in the Fe matrix show a similar trend along the series as the calculated spin moments, 
with a transition between antiferromagnetic and ferromagnetic coupling located between Os and Ir, and with magnetic moment values ranging from -0.4 to 0.5~$\mu_{\mathrm{B}}$\,atom$^{-1}$~\cite{wienke_determination_1991}.
The shift of the transition point between antiferromagnetic and ferromagnetic coupling, observed in our system between W and Re, in relation to the pure Fe matrix,  is due to the presence of Co, causing an increase in the number of electrons in our alloy and results from comparing the calculated spin magnetic moment with the measured total moment.
Simultaneously, the decrease of magnetic moment values in respect to the Fe matrix most probably results from the relatively high share of non-magnetic B atoms in the alloy, leading to the reduction of magnetic moments on Fe/Co atoms.
The discussed above results of measurements for 5$d$ impurities in the Fe matrix were initially theoretically predicted using the first-principles methods~\cite{akai_nuclear_1988, dederichs_ab-initio_1991}. 
Returning to the results shown in Fig.~\ref{fig:fecoxb_5d_mm}, 
we further notice that the function of orbital moment dependence on the type of 5$d$ impurity resembles the course of the sinus function with a minimum and maximum on the negative and positive side of the value range.
Very similar behavior for 5$d$ impurities in Fe matrix has been predicted previously, with the minimum for Os and maximum for Pt~\cite{dederichs_ab-initio_1991}.
We have also presented similar results for 5$d$ impurities in Fe$_5$PB$_2$ compound~\cite{werwinski_magnetocrystalline_2018}.

%
\begin{table}
\caption{
Spin ($m_\mathrm{s}$) and orbital  ($m_\mathrm{l}$) magnetic moments [$\mu_{\mathrm{B}}$ (atom or formula unit)$^{-1}$] of 
(Fe$_{0.66}$Co$_{0.28}$W$_{0.06}$)$_2$B,
(Fe$_{0.66}$Co$_{0.28}$Re$_{0.06}$)$_2$B, and
(Fe$_{0.7}$Co$_{0.3}$)$_2$B
as calculated for spin quantization axis along the $c$-axis.
The calculation was performed using fully-relativistic FPLO18 code with PBE functional.
%
\vspace{2mm}
}
\centering
\begin{tabular}{crrrrrr}
\hline \hline
		&\multicolumn{2}{c}{\scriptsize(Fe$_{0.66}$Co$_{0.28}$W$_{0.06}$)$_2$B}&\multicolumn{2}{c}{\scriptsize{(Fe$_{0.66}$Co$_{0.28}$Re$_{0.06}$)$_2$B}}&\multicolumn{2}{c}{\scriptsize(Fe$_{0.7}$Co$_{0.3}$)$_2$B}\\
\hline       
		&$m_\mathrm{s}$	&$m_\mathrm{l}$	&$m_\mathrm{s}$	&$m_\mathrm{l}$	&$m_\mathrm{s}$	&$m_\mathrm{l}$\\
\hline
Fe/Co	& 1.67			&0.04			&1.69			&0.05			&1.81			&0.04\\
B		&-0.18			&0.00			&-0.18			&0.00			&-0.22			&0.00\\
M		&-0.11			&-0.02			&0.05			&-0.05			&--				&--\\
total	&2.93			&0.08			&2.99			&0.08			&3.41			&0.09\\
\hline \hline
\label{tab:mm} 
\end{tabular}
\end{table}

Table~\ref{tab:mm} presents the spin ($m_\mathrm{s}$) and orbital ($m_\mathrm{l}$) magnetic moments of considered compositions: (Fe$_{0.66}$Co$_{0.28}$W$_{0.06}$)$_2$B, (Fe$_{0.66}$Co$_{0.28}$Re$_{0.06}$)$_2$B, and (Fe$_{0.7}$Co$_{0.3}$)$_2$B.
The $m_\mathrm{l}$ on the virtual Fe/Co atom is relatively low and underestimated by application of PBE approximation.
For (Fe$_{0.7}$Co$_{0.3}$)$_2$B the $m_\mathrm{s}$ on virtual Fe/Co atom is equal to 1.81~$\mu_{\mathrm{B}}$ and the induced $m_\mathrm{s}$ on B is equal to \mbox{-0.22}~$\mu_{\mathrm{B}}$.
The substitution of the considered alloy with W and Re leads to a reduction of several percent of  $m_\mathrm{s}$ on both virtual Fe/Co atom and B, whereas the small induced magnetic moments on W and Re have been discussed before.
In summary, the admixture of W and Re leads to a total moment per formula unit of 2.93~$\mu_{\mathrm{B}}$ for (Fe$_{0.66}$Co$_{0.28}$W$_{0.06}$)$_2$B and 2.99~$\mu_{\mathrm{B}}$ for (Fe$_{0.66}$Co$_{0.28}$Re$_{0.06}$)$_2$B, as compared to 3.41~$\mu_{\mathrm{B}}$ for (Fe$_{0.7}$Co$_{0.3}$)$_2$B.

%
%
\begin{figure}[t!]
\centering
\includegraphics[clip, width = 0.99 \columnwidth]{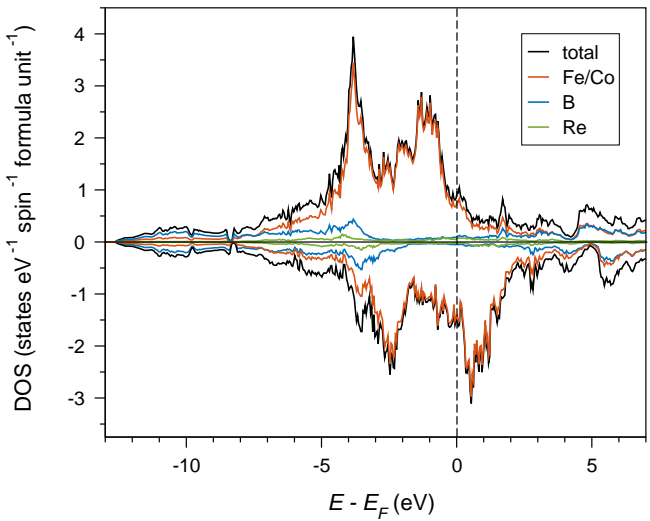}
\caption{
Densities of states (DOS) of (Fe$_{0.66}$Co$_{0.28}$Re$_{0.06}$)$_2$B alloy.
The model is combined of supercell and virtual crystal approximation (VCA).
The calculation was performed using FPLO18 code with PBE functional.
}
\label{fig:fecoreb_dos} 
\end{figure}
The discussed spin polarization is clearly visible on the plot of spin-polarized densities of states (DOS) of (Fe$_{0.66}$Co$_{0.28}$Re$_{0.06}$)$_2$B, see Fig.~\ref{fig:fecoreb_dos}.
The plot covers the range of the valence band of several eV's around the Fermi level.
The main contribution comes from Fe/Co 3$d$ orbitals.
Contribution from Re, although noticeably small, 
leads to changes in the band structure leading to the doubling of the MAE.
These changes are barely visible in this scale, but can be traced on a scale of tenths of eV around the Fermi level~\cite{edstrom_magnetic_2015}.

%
\begin{figure}[!t]
\centering
\includegraphics[clip, width=0.97\columnwidth]{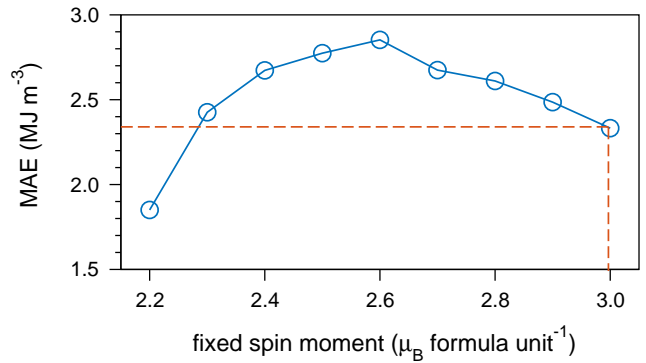}
\caption{
Magnetocrystalline anisotropy energy (MAE) of (Fe$_{0.66}$Co$_{0.28}$Re$_{0.06}$)$_2$B alloy as a function of the fixed spin magnetic moment.
The dashed line means MAE equal to 2.34~MJ\,m$^{-3}$ for equilibrium spin magnetic moment 2.99~$\mu_{\mathrm{B}}$\,f.u.$^{-1}$. 
The calculation was performed using FPLO18 code with PBE functional.
The model is combined of supercell and virtual crystal approximation (VCA).
}
\label{fig:fecoreb_fsm}
\end{figure}

The experiments have shown that the magnetic anisotropy constants of some (Fe,Co)$_2$B alloys change significantly with temperature~\cite{iga_magnetocrystalline_1970}.
However, for Co concentration \textit{x}~=~0.3 it behaves rather conventional, monotonically decreasing with temperature~\cite{iga_magnetocrystalline_1970}.
The question arises, whether this behavior will also be true for the (Fe$_{0.7}$Co$_{0.3}$)$_2$B sample doped with Re?
To answer it, we calculate how the MAE changes with the reduction of the spin magnetic moment.
Our conclusion is based on the expected relationship, that an increase in temperature causes a decrease in a magnetic moment~\cite{edstrom_magnetic_2015}.
In this way, by studying magnetic moment reduction, we gain insight into  the behavior of MAE with temperature.
In Fig.~\ref{fig:fecoreb_fsm} we present the plot of MAE as a function of the fixed spin magnetic moment for (Fe$_{0.66}$Co$_{0.28}$Re$_{0.06}$)$_2$B alloy.
As the magnetic moment decreases, the MAE first increases to about 2.9~MJ\,m$^{-3}$ and then begins to decrease, reaching its initial value of about 2.4~MJ\,m$^{-3}$ at about 25\% of the moment decrease ($m_s = 2.3$\,$\mu_{\mathrm{B}}$\,f.u.$^{-1}$).
As it has been shown experimentally for similar alloy (Fe$_{0.75}$Co$_{0.25}$)$_2$B~\cite{edstrom_magnetic_2015}, a decrease in magnetization of about 25\% occurs at about 700 K.
Thus we would expect that the predicted large MAE of (Fe$_{0.66}$Co$_{0.28}$Re$_{0.06}$)$_2$B will be stable up to relatively high temperatures.
However, thermal effects can also affect the electronic structure in other ways than just destabilizing magnetic interactions, leading to a subsequent decrease in MAE.

\subsection{\label{sec:level2}Structural characterization}

\begin{figure}[tb]
	\includegraphics[width=0.45\textwidth]{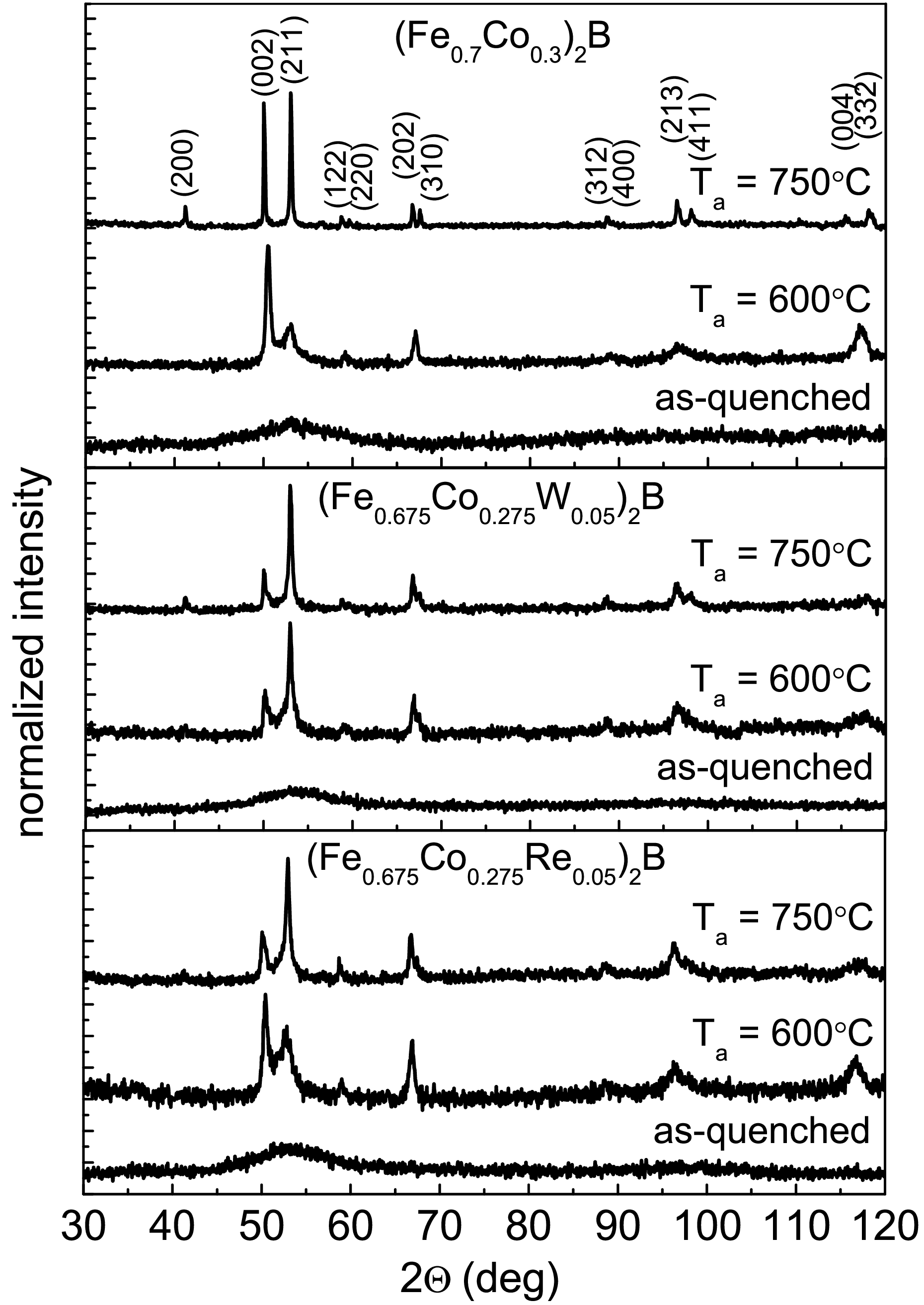}
	\caption{X-ray diffraction patterns of ~(Fe$_{0.7}$Co$_{0.3}$)$_2$B, ~(Fe$_{0.675}$Co$_{0.275}$W$_{0.05}$)$_2$B, and ~(Fe$_{0.675}$Co$_{0.275}$Re$_{0.05}$)$_2$B alloys in as-quenched state and annealed at \textit{T}$_a$ = 600 and 750$^o$C for 60 min. Miller indexes of CuAl$_2$-type structure are assigned.}
	\label{fig:XRD}
\end{figure}

X-ray diffraction measurements were performed mainly to characterize as-quenched samples, their transformation route during heat treatment, and to confirm single phase character of annealed samples. Results for as-quenched and annealed (Fe$_{0.7}$Co$_{0.3}$)$_2$B, (Fe$_{0.675}$Co$_{0.275}$W$_{0.05}$)$_2$B, and (Fe$_{0.675}$Co$_{0.275}$Re$_{0.05}$)$_2$B alloys are shown in Fig.~\ref{fig:XRD}. All samples are amorphous in the as-quenched state and crystallize into tetragonal (Fe,Co)$_2$B structure (CuAl$_2$-type with the \textit{I}4/\textit{mcm} space group) \cite{Barinov2007} after annealing. A similar situation has already been reported for Co-B systems \cite{Hasegawa1979}. The shortest interatomic distances \textit{R} are equal to 2.44, 2.42, and 2.45~Å for (Fe$_{0.7}$Co$_{0.3}$)$_2$B, (Fe$_{0.675}$Co$_{0.275}$W$_{0.05}$)$_2$B, and (Fe$_{0.675}$Co$_{0.275}$Re$_{0.05}$)$_2$B, respectively. These values are slightly larger than the average of interatomic distances between different atoms in the calculated FeCoB relaxed tetragonal structure, where \textit{R} is equal to 2.35~Å \cite{Khan2014}. Comparable results have been also obtained for amorphous alloy Fe$_8$Co$_{80}$Zr$_7$B$_4$Cu$_1$ (\textit{R} = 2.53~Å) \cite{McHenry1999a}. The position \textit{8h} of Co atoms in tetragonal Co$_2$B phase \cite{Rogl1979a} and Fe atoms in Fe$_2$B phase are slightly different.  With the increase of annealing temperature, some peaks (68$^o$, 97$^o$, and 116$^o$) split into two independent reflections. Moreover, alteration of the intensity ratio of (002) and (211) reflections is observed after isothermal annealing at 750$^o$C and already at 600$^o$C for (Fe$_{0.675}$Co$_{0.275}$W$_{0.05}$)$_2$B.  The (200) reflection appeared only after annealing at 750$^o$C and its intensity is relatively low. Lattice parameters were calculated by the use of the Rietveld method and are in agreement with values used for theoretical calculation and also similar to those reported before \cite{Jian2014,edstrom_magnetic_2015}. Lattice parameters strongly depend on the annealing temperature and  chemical composition of the alloys, as shown in Table \ref{tab:table1}. With the increasing annealing temperature, lattice parameter \textit{a} decreases and \textit{c} increases, as in Ref. \cite{iga_magnetocrystalline_1970}.

\begin{table}[htbp]
\caption{Lattice parameters and their ratio determined from X-ray diffraction analysis for (Fe$_{0.7}$Co$_{0.3}$)$_2$B, (Fe$_{0.675}$Co$_{0.275}$W$_{0.05}$)$_2$B, and (Fe$_{0.675}$Co$_{0.275}$Re$_{0.05}$)$_2$B alloys after isothermal annealing at annealing temperature \textit{T}$_a$ equal to 600 and 750$^o$C for 60 min.}
\begin{tabular}{cm{0.07\textwidth}<{\centering}m{0.06\textwidth}<{\centering}m{0.06\textwidth}<{\centering}m{0.06\textwidth}<{\centering}}
\hline \hline
alloy & \textit{T}$_a$ [$^o$C] & \textit{a} [Å] & \textit{c} [Å] & \textit{c/a} \\
\hline
\multirow{2}{*}{(Fe$_{0.7}$Co$_{0.3}$)$_2$B} & 600 & 5.128 & 4.192 & 0.817 \\
& 750 & 5.084 & 4.228 & 0.831 \\
\hline
\multirow{2}{*}{(Fe$_{0.675}$Co$_{0.275}$W$_{0.05}$)$_2$B} & 600 & 5.100 & 4.214 & 0.826 \\
& 750 & 5.092 & 4.224 & 0.829 \\
\hline
\multirow{2}{*}{(Fe$_{0.675}$Co$_{0.275}$Re$_{0.05}$)$_2$B} & 600 & 5.150 & 4.199 & 0.815 \\
& 750 & 5.107 & 4.222 & 0.826 \\
\hline \hline
\end{tabular}
  \label{tab:table1}
\end{table}

\subsection{\label{sec:level2}Thermal stability}

\begin{figure}[!h]
	\includegraphics[width=0.45\textwidth]{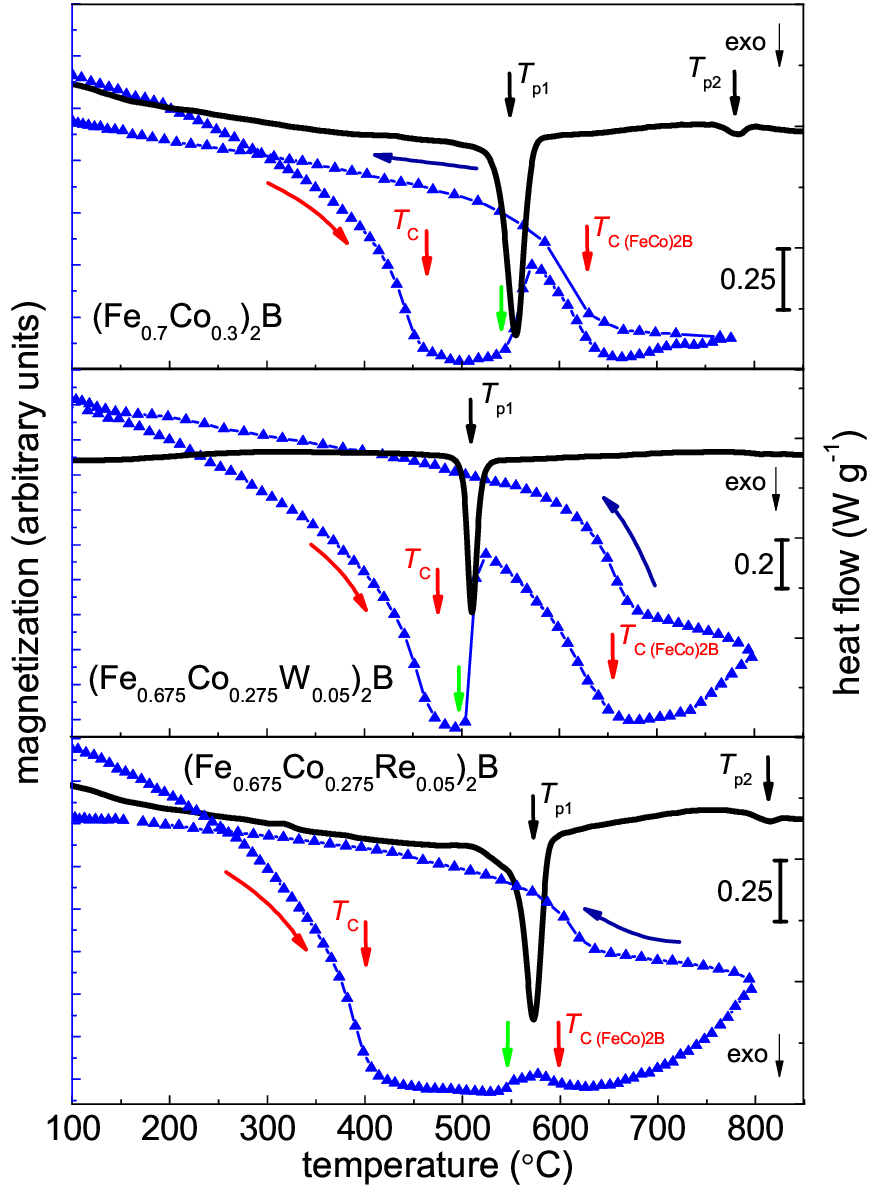}
	\caption{Thermomagnetic curves (blue points) set with differential scanning calorimetry traces (black line) for (Fe$_{0.7}$Co$_{0.3}$)$_2$B, (Fe$_{0.675}$Co$_{0.275}$W$_{0.05}$)$_2$B, and (Fe$_{0.675}$Co$_{0.275}$Re$_{0.05}$)$_2$B alloys in the as-quenched state, measured at heating and cooling rates of 10 K min$^{-1}$. Thermomagnetic curves were collected in a magnetic field of 2.5 kOe.}
	\label{fig:MTvsDSC}
\end{figure}

Calorimetric measurements were performed to determine thermal stability of amorphous phase and characteristic temperatures of crystallization of tetragonal phase, see Fig. \ref{fig:MTvsDSC}. For each case, one distinct exothermic reaction is observed between 500 and 600$^o$C. Small exothermic effects were also detected at around 800$^o$C for (Fe$_{0.7}$Co$_{0.3}$)$_2$B and (Fe$_{0.675}$Co$_{0.275}$Re$_{0.05}$)$_2$B. The onset temperature of crystallization (\textit{T}$_{onset}$) for (Fe$_{0.7}$Co$_{0.3}$)$_2$B sample is equal to 536$^o$C, with maxima of phase transitions for the first and the second process at \textit{T}$_{p1}$ = 551$^o$C and \textit{T}$_{p2}$ = 780$^o$C. Their enthalpies are equal to $\Delta\textit{H}_1 \approx$ 109 J g$^{-1}$ and $\Delta\textit{H}_2 \approx$ 4 J g$^{-1}$. \textit{T}$_{onset}$ for (Fe$_{0.675}$Co$_{0.275}$W$_{0.05}$)$_2$B decreases to 503$^o$C, indicating lower thermal stability than for the (Fe$_{0.7}$Co$_{0.3}$)$_2$B alloy (\textit{T}$_{p1}$ = 510$^o$C and $\Delta\textit{H}_1 \approx$ 121 J g$^{-1}$). In the case of Re substitution \textit{T}$_{onset}$ = 560$^o$C indicates higher thermal stability in comparison to (Fe$_{0.7}$Co$_{0.3}$)$_2$B. As mentioned above, two-phase transitions were detected in Re-doped alloy, where \textit{T}$_{p1}$ = 574$^o$C, \textit{T}$_{p2}$ = 815$^o$C, and $\Delta\textit{H}_1 \approx$ 112 J g$^{-1}$, $\Delta\textit{H}_2 \approx$ 3 J g$^{-1}$, respectively. In all the investigated alloys, the first crystallization peak occurs at higher temperatures than determined previously by Spassov and Diakovich for amorphous Fe$_{76}$Co$_{4}$B$_{20}$ alloy \cite{Spassov1993b}. It can be associated with higher boron content as has been shown by Fukamichi \textit{et al.} \cite{Fukamichi1977}. Authors have proposed that an increase of boron content in amorphous Fe-B alloys leads to enhancement of crystallization temperature. Differences in \textit{T}$_{onset}$ are governed by changing atomic packing, which in turn is related to the shortest interatomic distances. \textit{T}$_{onset}$ for (Fe$_{0.675}$Co$_{0.275}$W$_{0.05}$)$_2$B alloy is the lowest and coincides with the lowest calculated \textit{R} value. 

\begin{figure}[!h]
	\includegraphics[width=0.45\textwidth]{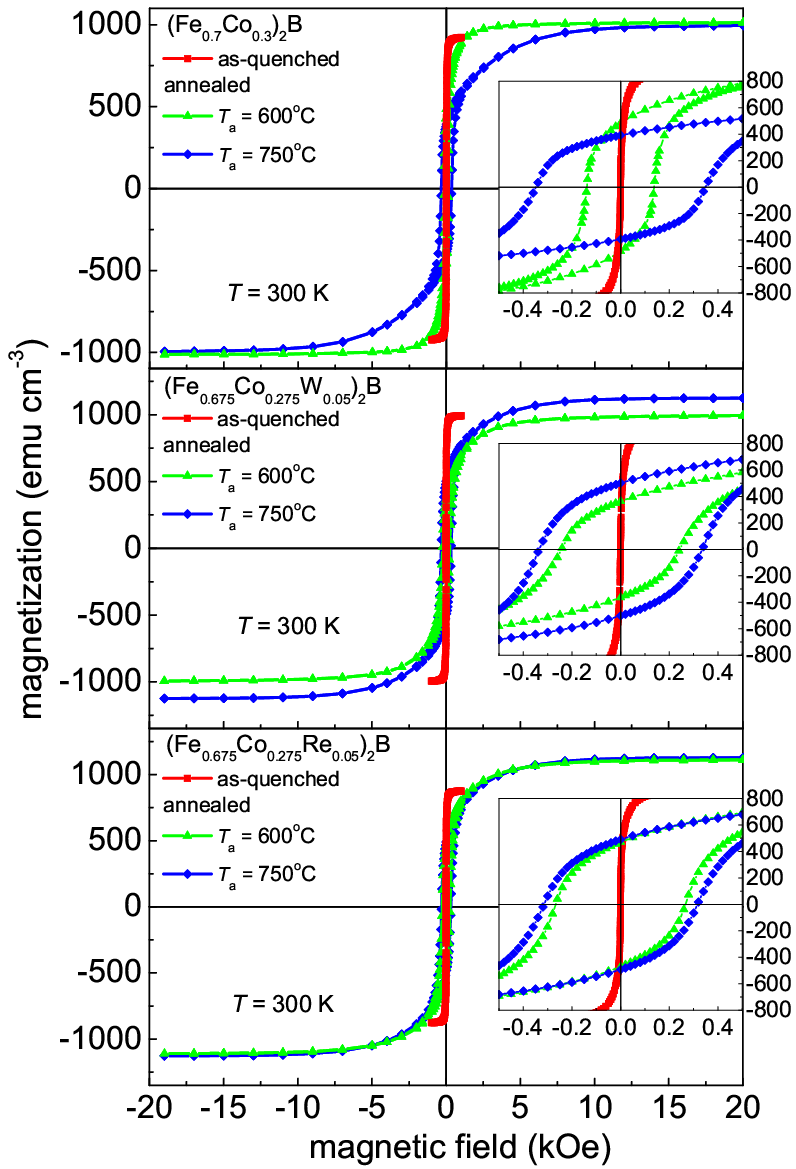}
	\caption{Magnetic hysteresis loops measured at room temperature for (Fe$_{0.7}$Co$_{0.3}$)$_2$B, (Fe$_{0.675}$Co$_{0.275}$W$_{0.05}$)$_2$B, and (Fe$_{0.675}$Co$_{0.275}$Re$_{0.05}$)$_2$B alloys in as-quenched state and after annealing at \textit{T}$_a$ equal to 600 and 750$^o$C for 60 min.}
	\label{fig:MH}
\end{figure}

\begin{table}[hbp]
\caption{Thermal properties as maxima of phase transitions for the first and the second process (\textit{T}$_{p1}$ and \textit{T}$_{p2}$), enthalpies of phase transition for the first and the second process $\Delta\textit{H}_1$ and $\Delta\textit{H}_2$ of (Fe$_{0.7}$Co$_{0.3}$)$_2$B, (Fe$_{0.675}$Co$_{0.275}$W$_{0.05}$)$_2$B, and (Fe$_{0.675}$Co$_{0.275}$Re$_{0.05}$)$_2$B alloys in as-quenched state.}
\begin{tabular}{cp{0.06\textwidth}<{\centering}p{0.06\textwidth}<{\centering}p{0.06\textwidth}<{\centering}p{0.06\textwidth}<{\centering}}
\hline \hline
alloy & \textit{T}$_{p1}$ [$^o$C] & \textit{T}$_{p2}$ [$^o$C] &  $\Delta\textit{H}_1$ \newline[J g$^{-1}$] & $\Delta\textit{H}_2$ \newline[J g$^{-1}$] \\
\hline
{(Fe$_{0.7}$Co$_{0.3}$)$_2$B} & 551 & 780 & 109 & 4 \\
\hline
{(Fe$_{0.675}$Co$_{0.275}$W$_{0.05}$)$_2$B} & 510 & - & 121 & - \\
\hline
{(Fe$_{0.675}$Co$_{0.275}$Re$_{0.05}$)$_2$B} & 574 & 815 & 112 & 3 \\

\hline \hline
\end{tabular}
  \label{tab:table3}
\end{table}

\subsection{\label{sec:level2}Magnetic properties}
The magnetic properties of as-quenched and annealed samples were measured especially to determine the influence of 5\textit{d} elements (to make a firm comparison with DFT results), especially for W-containing sample, which was synthesized in single phase tetragonal structure for the first time. Thermomangetic measurements (performed in an external magnetic field of 2.5~kOe, with a heating and cooling rate of 10 K min$^{-1}$) play a key role, when compared with X-ray diffraction and calorimetric data.
Thermomagnetic and differential scanning calorimetry curves for the amorphous alloys are shown in Fig. \ref{fig:MTvsDSC}. Curie temperatures for (Fe$_{0.7}$Co$_{0.3}$)$_2$B and (Fe$_{0.675}$Co$_{0.275}$W$_{0.05}$)$_2$B are equal to 444 and 456$^o$C, respectively. (Fe$_{0.675}$Co$_{0.275}$Re$_{0.05}$)$_2$B alloy has the lowest \textit{T}$_C$ = 391$^o$C. The determined \textit{T}$_C$ are higher than those reported for Fe-Co-B alloys \cite{Sundar2005}, which may be related to higher boron content \cite{Fukamichi1977}. Above \textit{T}$_C$, at about 550$^o$C, the thermomagnetic curve rises as a result of crystallization of (Fe,Co)$_2$B phase for (Fe$_{0.7}$Co$_{0.3}$)$_2$B alloy. Similar behavior is observed in the (Fe$_{0.675}$Co$_{0.275}$W$_{0.05}$)$_2$B ribbon, where magnetization rises at about 500$^o$C, but with a further decrease of magnetization to zero due to ferromagnetic-paramagnetic transition of (Fe,Co)$_2$B phase at 634$^o$C. The determined \textit{T}$_C$ is lower than 662$^o$C reported before for this compound \cite{Jian2014}. As expected, magnetization increases at \textit{T}$_C$ during cooling. The crystallization temperature of (Fe$_{0.675}$Co$_{0.275}$Re$_{0.05}$)$_2$B is equal to about 550$^o$C, which is close to Curie temperature for the (Fe,Co)$_2$B phase. Thus, the crystallization process induces only a slight increase in magnetization.

Since the ribbons in the as-quenched state reached magnetization saturation (\textit{M}$_s$) in a magnetic field of 0.5 kOe, the measurements for these alloys were performed up to the magnetic field of 20 kOe, see Fig.\ref{fig:MH}. \textit{M}$_s$ is equal to 922, 993, and 876 emu cm$^{-3}$ for (Fe$_{0.7}$Co$_{0.3}$)$_2$B, (Fe$_{0.675}$Co$_{0.275}$W$_{0.05}$)$_2$B, and (Fe$_{0.675}$Co$_{0.275}$Re$_{0.05}$)$_2$B, respectively. Magnetization saturation is reached at about 15 kOe for all three ribbons after annealing and is higher than for the as-quenched samples. For alloys with W and Re substitutions, magnetization saturation increases with isothermal annealing temperature and for \textit{T}$_a$ = 750$^o$C is equal to 1124 and 1126 emu cm$^{-3}$ for alloys with W and Re atoms, respectively. In the case of (Fe$_{0.7}$Co$_{0.3}$)$_2$B alloy, \textit{M}$_s$ after isothermal annealing at 750$^o$C is equal to 995 emu cm$^{-3}$ and is lower than for the sample annealed at 600$^o$C (\textit{M}$_s$ = 1012 emu cm$^{-3}$). The value of magnetic moment is higher for the substituted samples than for the parent (Fe$_{0.7}$Co$_{0.3}$)$_2$B alloy, see Table \ref{tab:table4}, which contradicts our calculations in which we observed inverse relation. 

\begin{table}[htbp]
\caption{Magnetic properties such as Curie temperature (\textit{T}$_C$),
saturation magnetization (\textit{M}$_s$), coercive field (\textit{H}$_c$), and magnetic moment per formula unit (\textit{m}) of (Fe$_{0.7}$Co$_{0.3}$)$_2$B, (Fe$_{0.675}$Co$_{0.275}$W$_{0.05}$)$_2$B, and (Fe$_{0.675}$Co$_{0.275}$Re$_{0.05}$)$_2$B alloys in as-quenched state and after isothermal annealing at temperatures (\textit{T}$_a$) equal to 600 and 750$^o$C for 60 min.}
\begin{tabular}{cp{0.08\textwidth}<{\centering}p{0.025\textwidth}<{\centering}p{0.06\textwidth}<{\centering}p{0.03\textwidth}<{\centering}p{0.05\textwidth}<{\centering}}
\hline \hline
alloy & \textit{T}$_a$ \newline[$^o$C] & \textit{T}$_{C}$ [$^o$C] & \textit{M}$_{s}$ [emu cm$^{-1}$] &  \textit{H}$_c$ \newline[Oe] &$m$ \newline[$\mu{\textsubscript{B}}$ f.u.$^{-1}$] \\
\hline
\multirow{3}{*}{(Fe$_{0.7}$Co$_{0.3}$)$_2$B} & as-quenched & 458 & 922 & 10 & \\
& 600 & & 1012 & 138 & 2.99 \\ & 750 & &  995 & 348 & 2.94 \\
\hline
\multirow{3}{*}{(Fe$_{0.675}$Co$_{0.275}$W$_{0.05}$)$_2$B} & as-quenched & 456 & 993 & 10 & \\
& 600 & & 994 & 243 & 3.06 \\ & 750 & & 1124 & 335 & 3.46 \\
\hline
\multirow{3}{*}{(Fe$_{0.675}$Co$_{0.275}$Re$_{0.05}$)$_2$B} & as-quenched & 391 & 876 & 10 & \\
& 600 & & 1111 & 267 & 3.39 \\ & 750 & & 1126 & 315 & 3.44 \\

\hline \hline
\end{tabular}
  \label{tab:table4}
\end{table}
This discrepancy can be connected with samples’ microstructure, crystallinity, crystallite size, morphology. These factors are not taken into account in conducted calculations. According to this additional scanning electron microscopy (SEM) and energy-dispersive X-ray spectroscopy (EDX) measurements were made for the alloys annealed at 750°C (see Supplementary Fig. S1-S3). SEM micrographs and elemental maps show quite uniform microstructure and uniform distribution of elements in a microscale. 
Moreover, Williamson-Hall analysis was performed to determine changes in average crystallite size and lattice strain in annealed samples (see Table \ref{tab:table5}. Due to rather large uncertainties we are able to comment on some general trends. At first, the values of lattice strain are similar for analysed samples. Therefore, narrowing of X-ray diffraction peaks after annealing at 750°C can be ascribed to the increase of average crystallite size mostly. Even though the onset of crystallization of (Fe$_{0.7}$Co$_{0.3}$)$_2$B is higher than that determined for (Fe$_{0.675}$Co$_{0.275}$W$_{0.05}$)$_2$B and similar to the onset temperature of (Fe$_{0.675}$Co$_{0.275}$Re$_{0.05}$)$_2$B, we observe significant increase of average crystallite size for the parent alloy. It suggests that the presence of Re and W restrains the growth of crystallites. Such effect and differences in crystallite size and morphology of annealed samples are surely responsible for observed discrepancies in magnetic properties 
A low coercive fields (\textit{H}$_c$), on the order of a few Oe, were obtained for as-quenched alloys, which is typical for amorphous materials. Isothermal annealing at 600$^o$C leads to an increase of coercive field values up to 138, 243, and 267 Oe for (Fe$_{0.7}$Co$_{0.3}$)$_2$B, (Fe$_{0.675}$Co$_{0.275}$W$_{0.05}$)$_2$B, and (Fe$_{0.675}$Co$_{0.275}$Re$_{0.05}$)$_2$B, respectively. \textit{H}$_c$'s for substituted samples confirmed previous theoretical results obtained by Edstr\"om and coworkers \cite{edstrom_magnetic_2015}. Annealing at \textit{T}$_a$ = 750$^o$C improves the coercive field to the range of 315 - 348 Oe. The values obtained are still below the threshold of permanent magnets applications, but one has to bear in mind that these values can be maximized by further processing e.g. by magnetic field annealing or combined high pressure torsion and heat treatment \cite{Musia2019}.

\begin{table}[htbp]
\caption{The average crystallite size (\textit{D}) and lattice strain ($\epsilon$) determined on the basis of Williamson-Hall method for (Fe$_{0.7}$Co$_{0.3}$)$_2$B, (Fe$_{0.675}$Co$_{0.275}$W$_{0.05}$)$_2$B, and (Fe$_{0.675}$Co$_{0.275}$Re$_{0.05}$)$_2$B alloys after isothermal annealing at annealing temperature \textit{T}$_a$ equal to 600 and 750$^o$C for 60 min. \cite{WILLIAMSON195322}}.
\begin{tabular}{cm{0.07\textwidth}<{\centering}m{0.06\textwidth}<{\centering}m{0.07\textwidth}<{\centering}}
\hline \hline
alloy & \textit{T}$_a$ [$^o$C] & \textit{D} [nm] & \textit{$\epsilon$} x10$^{-3}$ \\
\hline
\multirow{2}{*}{(Fe$_{0.7}$Co$_{0.3}$)$_2$B} & 600 & 30$\pm$25 & 2.2$\pm$2.0 \\
& 750 & 150$\pm$61 & 1.1$\pm$0.2 \\
\hline
\multirow{2}{*}{(Fe$_{0.675}$Co$_{0.275}$W$_{0.05}$)$_2$B} & 600 & 32$\pm$17 & 1.5$\pm$1.4 \\
& 750 & 45$\pm$29 & 1.3$\pm$1.2 \\
\hline
\multirow{2}{*}{(Fe$_{0.675}$Co$_{0.275}$Re$_{0.05}$)$_2$B} & 600 & 3.6$\pm$2.8 & 30$\pm$25 \\
& 750 & 24$\pm$17 & 2.6$\pm$2.2 \\
\hline \hline
\end{tabular}
  \label{tab:table5}
\end{table}

\subsection{\label{sec:level2}M\"{o}ssbauer spectroscopy}

\subsubsection{Transmission M\"{o}ssbauer spectroscopy}

\begin{figure}[!ht]
	\includegraphics[width=0.5\textwidth]{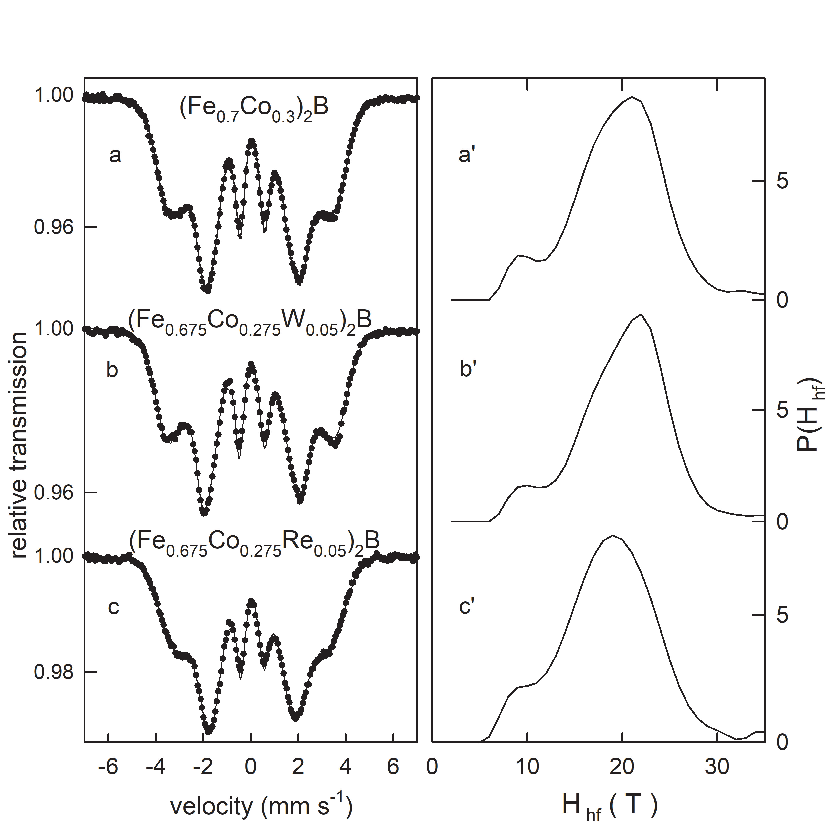}
	\caption{M\"{o}ssbauer spectra  and hyperfine field (\textit{H}$_{hf}$) distributions of (a) and (a') (Fe$_{0.7}$Co$_{0.3}$)$_2$B, (b) and (b') (Fe$_{0.675}$Co$_{0.275}$W$_{0.05}$)$_2$B, and (c), (c') (Fe$_{0.675}$Co$_{0.275}$Re$_{0.05}$)$_2$B alloys in as-quenched state.}
	\label{fig:widmo_mossbauer}
\end{figure}

Transmission and rf-Moessbauer spectroscopy studies were performed to confirm conclusions drawn on the basis of structural and magnetic measurements and to determine microscopic mechanisms underlying the effect of 5\textit{d} elements substitution. M\"{o}ssbauer spectra shown in Fig. \ref{fig:widmo_mossbauer} confirmed the amorphous structure for all three alloys in the as-quenched state. The spectra reveal unresolved magnetic hyperfine structures due to the lack of long-range order. The diversity of atomic environments around Fe nuclei in the amorphous phase is clearly visible in the broad hyperfine field distributions extracted from the spectra, see Fig. \ref{fig:widmo_mossbauer}. The dominating broad peak at about 20 T is accompanied by a small one at a reduced hyperfine field of about 9 T, related to minor Fe-depleted regions. The average magnetic hyperfine field determined for the as-quenched (Fe$_{0.7}$Co$_{0.3}$)$_2$B alloy is 20.1 T. For the alloy with W substitution this value is somewhat larger (20.6 T), while for the alloy with Re it is smaller and equals about 19 T. A similar trend is observed for saturation magnetization, see Table \ref{tab:table4}. The value of \textit{M}$_s$ was significantly higher for (Fe$_{0.675}$Co$_{0.275}$W$_{0.05}$)$_2$B as compared with the non-substituted and Re-substituted alloys.

\begin{figure}[!ht]
	\includegraphics[width=0.45\textwidth]{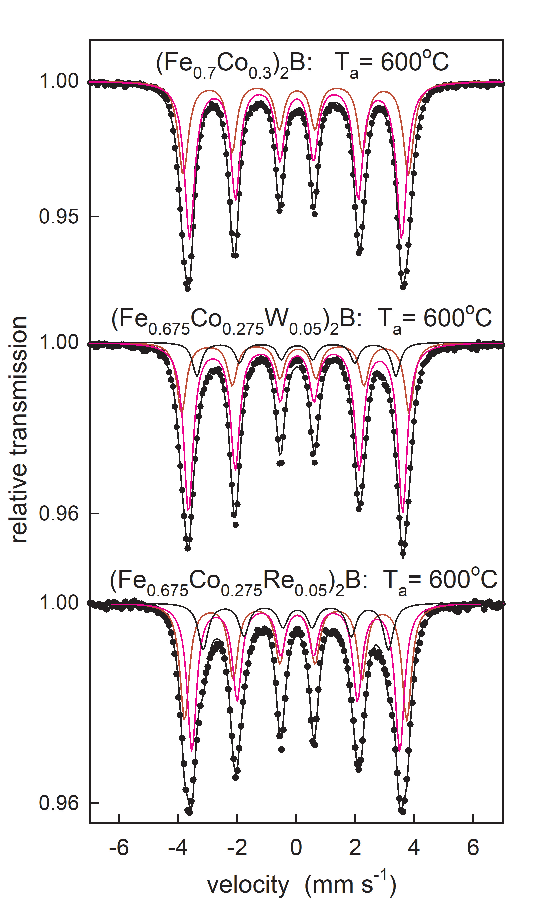}
	\caption{M\"{o}ssbauer spectra of (Fe$_{0.7}$Co$_{0.3}$)$_2$B, (Fe$_{0.675}$Co$_{0.275}$W$_{0.05}$)$_2$B, and (Fe$_{0.675}$Co$_{0.275}$Re$_{0.05}$)$_2$B alloys after annealing at \textit{T}$_a$ = 600$^o$C for 60 min.}
	\label{fig:mossbauer_annealed600}
\end{figure}

\begin{figure}[!ht]
	\includegraphics[width=0.45\textwidth]{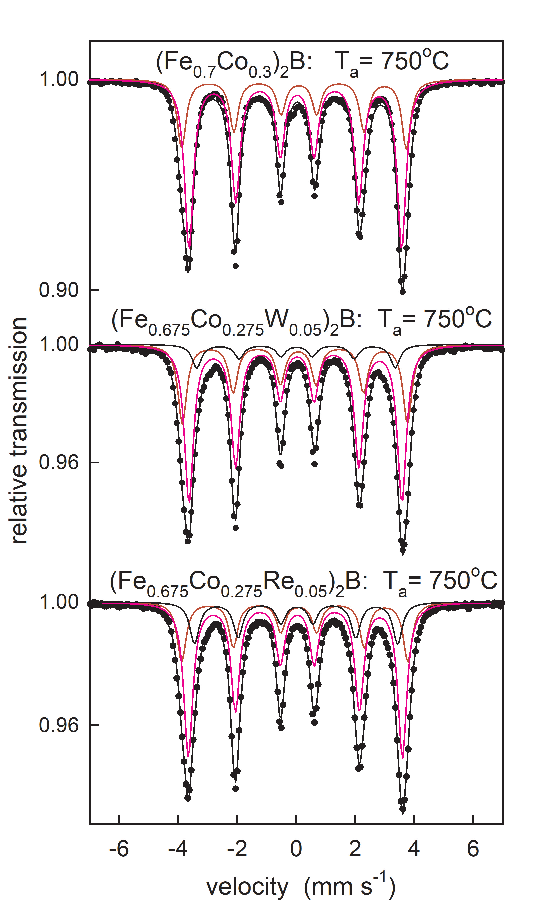}
	\caption{M\"{o}ssbauer spectra of (Fe$_{0.7}$Co$_{0.3}$)$_2$B, (Fe$_{0.675}$Co$_{0.275}$W$_{0.05}$)$_2$B, and (Fe$_{0.675}$Co$_{0.275}$Re$_{0.05}$)$_2$B alloys after annealing at \textit{T}$_a$ = 750$^o$C for 60 min.}
	\label{fig:mossbauer_annealed750}
\end{figure}

M\"{o}ssbauer spectra obtained for alloys annealed at 600$^o$C and 750$^o$C showed a complete crystallization of the amorphous phase. The spectra collected for the alloys after annealing at 600$^o$C for 60 min are shown in Fig. \ref{fig:mossbauer_annealed600}. 
The M\"{o}ssbauer spectrum for the (Fe$_{0.7}$Co$_{0.3}$)$_2$B alloy consists of two sextets, which is due to the Fe atoms occupying two magnetically inequivalent positions in the Fe$_2$B tetragonal structure. These sextets have slightly broadened lines due to incomplete crystalline order. The parameters of the hyperfine interactions related to these sextets are as follows: (1) hyperfine field  \textit{H}$_{hf1}$ = 23.7 T, isomer shift \textit{IS}$_1$ = +0.12 mm s$^{-1}$, the relative spectral fraction \textit{A}$_1$ = 38\% and (2) \textit{H}$_{hf2}$ = 22.3 T, \textit{IS}$_2$ = +0.11 mm s$^{-1}$, \textit{A}$_2$ = 62\%. Typical hyperfine parameters for the Fe$_2$B crystal phase are \textit{H}$_{hf1}$ = 24.0 T, \textit{H}$_{hf2}$ = 23.2 T, and the isomer shift of +0.17 mm s$^{-1}$ for both spectral components \cite{Kenneth1973, Weisman1969}. Thus, substitution of Co for Fe reduces the value of the hyperfine field in the alloy, which was also observed for the (Fe,Co)$_2$B phase formed in the (Fe,Co)-Pt-B alloys \cite{Grabias2017}. 
For the substituted alloys, three sextets were fitted in the M\"{o}ssbauer spectra, see Fig. \ref{fig:mossbauer_annealed600}. Parameters of the hyperfine interactions in all three alloys are presented in Table \ref{tab:table2}. For the first two sextets, the hyperfine parameters are similar to those for (Fe$_{0.7}$Co$_{0.3}$)$_2$B. The appearance of a third sextet with a significantly reduced hyperfine field (20-21 T) appears to be the result of substitution of Re or W elements that form an additional defect position.
The M\"{o}ssbauer spectra obtained after annealing at 750$^o$C, see Fig. \ref{fig:mossbauer_annealed750}, were fitted in the same way as the spectra for 600$^o$C. The hyperfine parameters of all three sextets are generally similar to those determined after annealing at 600$^o$C, see Table \ref{tab:table2}. However, the width of the sextet lines became narrower, indicating a well-crystallized structure after annealing at 750$^o$C. The hyperfine parameters of (Fe$_{0.675}$Co$_{0.275}$W$_{0.05}$)$_2$B alloy annealed at 600$^o$C are comparable to those determined for the sample annealed at 750$^o$C. It is worth noting that such consistency was also observed for lattice parameter, see Table \ref{tab:table1}. 
This is the effect of the lowest crystallization temperature (T$_{p1}$ = 510$^o$C) among all studied systems (Fig. \ref{fig:MTvsDSC}). This indicates that the (Fe$_{0.675}$Co$_{0.275}$W$_{0.05}$)$_2$B alloy is in fully crystallized and ordered form at the lowest annealing temperature already.
For the other two alloys, changes in the relative fractions of the subspectra are related to atomic ordering as the annealing temperature increases up to 750$^o$C, see Table \ref{tab:table2}. Based on the analysis of the Mössbauer spectra of all samples annealed at 750$^o$C, it appears that the additional substitution-related sextet (\textit{A}$_3$) partially replaces both main spectral components (\textit{A}$_1$ and \textit{A}$_2$). A substantial difference is observed in the relative spectral contribution of this additional component, which is significantly larger for the Re-substituted alloy than for the W-substituted one.

\begin{table*}[htbp]
\begin{center}
\begin{tabular}{  c  p{1.5cm}<{\centering} p{1.5cm}<{\centering}  p{2cm}<{\centering} p{2cm}<{\centering} p{2cm}<{\centering} p{2cm}<{\centering} }
\hline \hline
 & \multicolumn{2}{c}{\textbf{(Fe$_{0.7}$Co$_{0.3}$)$_2$B}} & \multicolumn{2}{c}{\textbf{(Fe$_{0.675}$Co$_{0.275}$Re$_{0.05}$)$_2$B}} & \multicolumn{2}{c}{\textbf{(Fe$_{0.675}$Co$_{0.275}$W$_{0.05}$)$_2$B}} \\ 
 \hline 
 \textbf{\textit{T}$_a$ [$^o$C]} & 600 & 750 & 600 & 750 & 600 & 750 \\ \hline     
 \textbf{\textit{H}$_h$$_f$$_1$ [T]} & 23.7 & 23.7 & 23.4 & 23.8 & 24.0 & 23.8 \\ 
 \textbf{\textit{H}$_h$$_f$$_2$ [T]} & 22.3 & 22.4 & 21.9 & 22.5 & 22.6 & 22.5 \\ 
 \textbf{\textit{H}$_h$$_f$$_3$ [T]} & - & - & 19.5 & 21.4 & 21.2 & 21.1 \\ \hline
 \textbf{\textit{IS}$_1$ [mm s$^{-1}$]} & 0.12 & 0.12 & 0.12 & 0.13 & 0.12 & 0.12 \\
 \textbf{\textit{IS}$_2$ [mm s$^{-1}$]} & 0.11 & 0.11 & 0.12 & 0.12 & 0.11 & 0.11 \\ 
 \textbf{\textit{IS}$_3$ [mm s$^{-1}$]} & - & - & 0.12 & 0.12 & 0.11 & 0.11 \\ \hline
 \textbf{\textit{QS}$_1$ [mm s$^{-1}$]} & -0.04 & -0.16 & -0.06 & -0.13 & -0.10 & -0.12 \\
 \textbf{\textit{QS}$_2$ [mm s$^{-1}$]} & -0.04 & -0.06 & -0.05 & -0.06 & -0.05 & -0.06 \\ 
 \textbf{\textit{QS}$_3$ [mm s$^{-1}$]} & - & - & -0.06 & -0.03 & -0.03 & -0.03 \\ \hline
 \textbf{\textit{A}$_1$ [\%]} & 38 & 29 & 38 & 24 & 25 & 25 \\
 \textbf{\textit{A}$_2$ [\%]} & 62 & 71 & 45 & 58 & 62 & 64 \\ 
 \textbf{\textit{A}$_3$ [\%]} & - & - & 17 & 18 & 13 & 11 \\  
 \hline \hline
\end{tabular}
\caption{The values of hyperfine field (\textit{H}$_{hf}$), isomer shift (\textit{IS}), quadrupole shift (\textit{QS}), and relative spectral area of each sextet (\textit{A}) for (Fe$_{0.7}$Co$_{0.3}$)$_2$B, (Fe$_{0.675}$Co$_{0.275}$W$_{0.05}$)$_2$B, and (Fe$_{0.675}$Co$_{0.275}$Re$_{0.05}$)$_2$B alloys after annealing at 600 and 750$^o$C for 60 min.}
\label{tab:table2}
\end{center}
\end{table*}

\subsubsection{rf-M\"{o}ssbauer spcectroscopy}

\begin{figure}[h!]
	\includegraphics[width=0.5\textwidth]{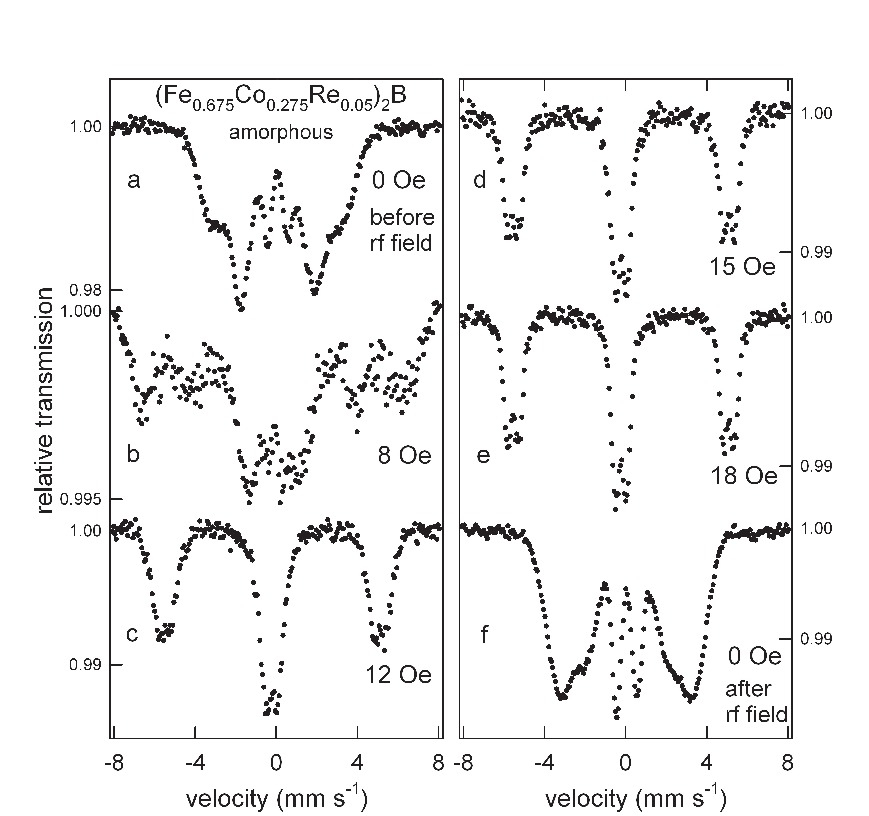}
	\caption{rf-M\"{o}ssbauer spectra of the amorphous (Fe$_{0.675}$Co$_{0.275}$Re$_{0.05}$)$_2$B alloy in various rf-fields.}
	\label{fig:rf-mossbauer_FeCoReB}
\end{figure}

\begin{figure}[h!]
	\includegraphics[width=0.5\textwidth]{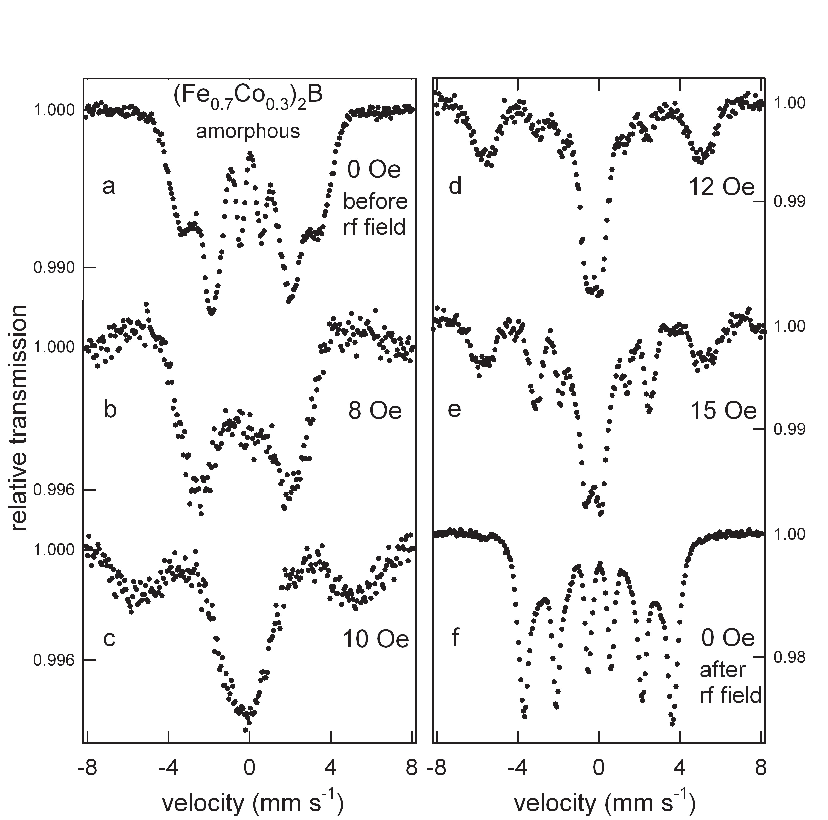}
	\caption{rf-M\"{o}ssbauer spectra of the amorphous (Fe$_{0.7}$Co$_{0.3}$)$_2$B alloy in various rf-fields.}
	\label{fig:rf-mossbauer_FeCoB}
\end{figure}

\begin{figure}[!h]
	\includegraphics[width=0.5\textwidth]{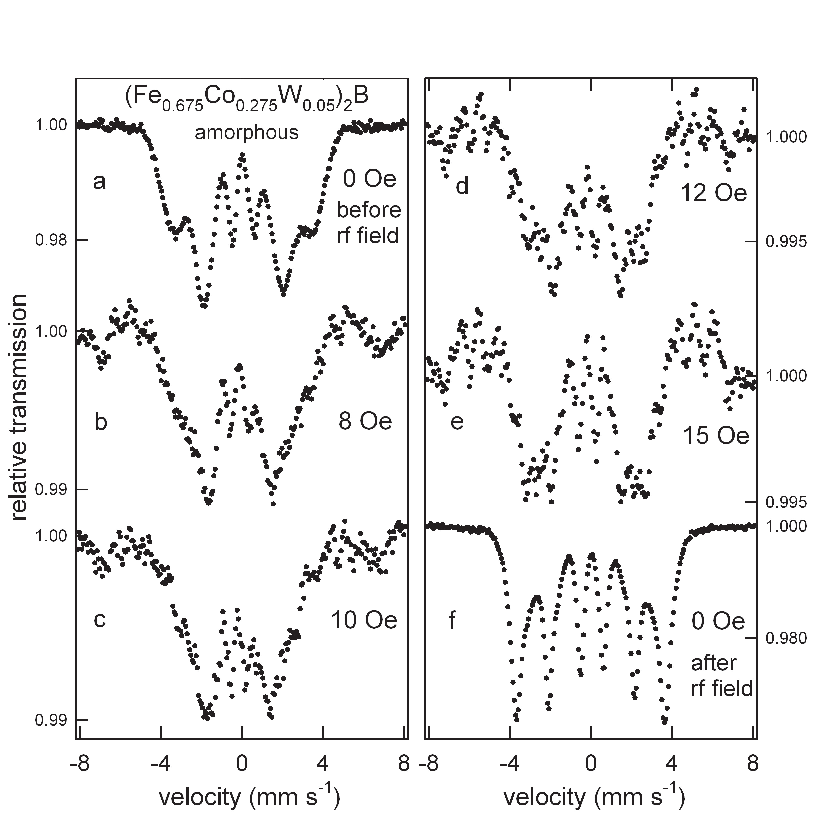}
	\caption{rf-M\"{o}ssbauer spectra of the amorphous (Fe$_{0.675}$Co$_{0.275}$W$_{0.05}$)$_2$B alloy in various rf-fields.}
	\label{fig:rf-mossbauer_FeCoWB}
\end{figure}

The influence of chemical composition in the amorphous alloys was studied by the unconventional rf-M\"{o}ssbauer technique described in detail elsewhere \cite{Kopcewicz1991}. 
As the most representative example, the Mössbauer spectra obtained for the (Fe$_{0.675}$Co$_{0.275}$Re$_{0.05}$)$_2$B alloy are shown in Fig. \ref{fig:rf-mossbauer_FeCoReB}. 
The rf-M\"{o}ssbauer spectra reveal that the amorphous alloy is magnetically soft with a low coercive field in the range of a few Oe as evidenced by significant narrowing of the magnetically split sextet, which is clearly seen already at about 8 Oe, see Fig. \ref{fig:rf-mossbauer_FeCoReB}(b). As the radio-frequency field intensity increases, a complete collapse of the magnetic hyperfine structure into a non-magnetic quadrupole doublet is observed, see Fig. \ref{fig:rf-mossbauer_FeCoReB}(c)-\ref{fig:rf-mossbauer_FeCoReB}(e). The rf-collapse effect is due to the fast magnetization reversal, induced by the rf field, which leads to averaging of the hyperfine field to zero \cite{Kopcewicz1991}. This means that the intensity of the applied rf-field is large enough to overcome the local magnetic anisotropy field. The rf field also has another effect whereby sidebands appear in the spectrum \cite{Kopcewicz1991}. The rf sidebands visible at about +/- 5 mm s$^{-1}$ reflect the shape of the collapsed spectrum, i.e. quadrupole doublets, see Fig. \ref{fig:rf-mossbauer_FeCoReB}(c)-\ref{fig:rf-mossbauer_FeCoReB}(e). Their high intensity indicates that the alloy has large magnetostriction. The spectrum recorded after exposure to the rf field shows that the sample has retained an amorphous structure, see Fig. \ref{fig:rf-mossbauer_FeCoReB}(a), \ref{fig:rf-mossbauer_FeCoReB}(f). The average hyperfine field is 18.9 T, which is actually identical to that of the as-quenched alloy. The slightly different shape of the spectrum obtained after rf field exposure in comparison with the spectrum of the as-quenched sample is related to a significant reduction in the intensity of the second and fifth lines. This indicates a change from a preferential in-plane spin alignment for the as-quenched alloy to a more out-of-plane arrangement after rf field exposure. The most probable reason for this feature is related to surface stresses induced by the formation of some crystallites on the surface of the ribbon (surface crystallization), while the bulk of the ribbon remains amorphous.

A similar dependence of the rf-Mössbauer spectra was recorded for the (Fe$_{0.7}$Co$_{0.3}$)$_2$B amorphous alloy, see Fig. \ref{fig:rf-mossbauer_FeCoB}. A narrowing of the magnetic hyperfine structure is observed at 10 Oe, while the complete rf-collapse effect occurs at the rf field intensity of 12 Oe, see Fig. \ref{fig:rf-mossbauer_FeCoB}(b)-\ref{fig:rf-mossbauer_FeCoB}(d). The rf sidebands are significantly less intense than those observed for Re substitution, strongly suggesting smaller magnetostriction of the non-substituted alloy. In contrast to the (Fe$_{0.675}$Co$_{0.275}$Re$_{0.05}$)$_2$B ribbon, which remained fully amorphous after exposure to the rf field, the (Fe$_{0.7}$Co$_{0.3}$)$_2$B alloy starts to crystallize at the intensity of about 12 Oe. The Mössbauer spectrum recorded after exposure to the rf field is completely different from the spectrum of the as-quenched alloy, see Fig. \ref{fig:rf-mossbauer_FeCoB}(a) and \ref{fig:rf-mossbauer_FeCoB}(f), but similar to that of the annealed sample (Fig. \ref{fig:mossbauer_annealed600}). This indicates that the amorphous phase present in the as-quenched alloy was almost completely crystallized when the sample was exposed to the rf field of 15 Oe. For the (Fe$_{0.675}$Co$_{0.275}$W$_{0.05}$)$_2$B alloy, the narrowing of the magnetically split sextet is barely visible at 10 Oe and higher rf fields (Fig. \ref{fig:rf-mossbauer_FeCoWB}). The small rf sidebands observed for this sample are evidence of non-zero magnetostriction. Furthermore, similarly to the case of the non-substituted alloy, the crystallization process starts at the rf field intensity of about 12 Oe, see Fig. \ref{fig:rf-mossbauer_FeCoWB}(d), \ref{fig:rf-mossbauer_FeCoWB}(e). The spectrum obtained after rf field exposure, see Fig. \ref{fig:rf-mossbauer_FeCoWB}(f), resembles the corresponding spectrum of the undoped alloy shown in Fig. \ref{fig:rf-mossbauer_FeCoB}(f). Such rf field-induced crystallization, which was observed in the case of (Fe$_{0.7}$Co$_{0.3}$)$_2$B and (Fe$_{0.675}$Co$_{0.275}$W$_{0.05}$)$_2$B amorphous alloys, has been previously reported for Co-containing amorphous alloys such as FeCoZrSi \cite{Kopcewicz2011}. The origin of rf-crystallization is most probably related to mechanical deformations forced in the samples by the rf field via magnetoacoustic vibrations \cite{Kopcewicz2011}. It seems that among the FeCoB alloys studied, the addition of Re plays a crucial role in the stabilization of the amorphous phase.

\section{Conclusions}

Density functional theory calculations and complementary experimental methods, as for example X-ray diffraction, magnetometry, and M\"{o}ssbauer spectroscopy techniques were utilized to determine physical properties of (Fe,Co)$_2$B alloys with specific 5\textit{d} substitutions. 
The main objective of the undertaken research was to evaluate their suitability as hard magnetic materials. 

%
The first-principles calculation were applied to the systematic study of (Fe,Co)$_2$B alloys with 3$d$, 4$d$, and 5$d$ substitutions.
Their results show a twofold increase in magnetocrystalline anisotropy for (Fe$_{0.66}$Co$_{0.28}$Re$_{0.06}$)$_2$, from 1.17 to 2.34~MJ\,m$^{-3}$. 
However, the determined values are significantly overestimated due to the use of the virtual crystal approximation. 
Calculations also show that no other dopants from the $d$-block introduce a significant improvement in magnetocrystalline anisotropy. 
In contrast to previous results, this means that no noticeable increase in magnetocrystalline anisotropy is predicted for the W-doped system.
Calculations also indicate relatively good temperature stability of the magnetocrystalline anisotropy of Re-doped alloy in the relevant temperature range up to 700~K. 
When (Fe$_{0.7}$Co$_{0.3}$)$_2$B is doped with both Re or W, the total magnetization decreases by 12-14\%.

%
(Fe$_{0.7}$Co$_{0.3}$)$_2$B, (Fe$_{0.675}$Co$_{0.275}$Re$_{0.05}$)$_2$B, and (Fe$_{0.675}$Co$_{0.275}$W$_{0.05}$)$_2$B precursors were synthesized by melt-spinning method. 
After annealing, all of the mentioned samples crystallized in a desired tetragonal (Fe,Co)$_2$B structure (CuAl$_2$-type with space group \textit{I}4/\textit{mcm}). 
With increasing annealing temperature, we observed a decrease in \textit{a} and an increase in \textit{c} lattice parameter.
Calorimetric measurements exhibit the presence of significant exothermic reaction in the range of 500 - 600$^o$C for all the alloys considered.
In addition, small exothermic effects were also detected at around 800$^o$C for (Fe$_{0.7}$Co$_{0.3}$)$_2$B and (Fe$_{0.675}$Co$_{0.275}$Re$_{0.05}$)$_2$B. 
Changes in the onset temperature are correlated with atomic packing and the shortest interatomic distances. 
The more densely packed structure (with shorter interatomic distances), the higher the value of the onset temperature and the higher is the thermal stability of amorphous phase. 
Due to this fact, the onset temperature for the alloy (Fe$_{0.675}$Co$_{0.275}$W$_{0.05}$)$_2$B is the lowest and coincides with the lowest estimated value of the shortest interatomic distance. 
Thermomagnetic measurements confirmed the crystallization of the (Fe,Co)$_2$B phase as evidenced by a decrease in magnetization at about 600$^o$C (\textit{T}$_C$ of the (Fe,Co)$_2$B alloy).
%
%
The value of magnetic moment is higher for the substituted samples than for the parent (Fe$_{0.7}$Co$_{0.3}$)$_2$B alloy,
see Table IV, which contradicts our calculations in which we observed the inverse relation. At first, calculations do not take into account any extrinsic parameters, as for example those connected with samples microstructure. Secondly, even though all the alloys were isothermally annealed at the same temperatures, such treatment leads to crystallization products which differ by a quality of the crystalline structure for each composition due to variations in crystallization temperatures. So one has to bear in mind that the influence of these factors on the value of magnetic moment/magnetization for particular alloy, can have a bigger impact than simply the intrinsic effect of the W or Re substitution.
Moreover, coercivity increased to above 300~Oe in all of the considered samples after crystallization of (Fe,Co)$_2$B phase, which \textit{T}$_C$ is just slightly dependent on 5$d$ substitution. 
More importantly, both Re and W improve saturation magnetization of heat treated samples in comparison to parent (Fe$_{0.7}$Co$_{0.3}$)$_2$B alloy.

%
Mössbauer spectroscopy measurements showed that the crystallization process of the amorphous alloys was completed after annealing at 750$^o$C. 
In addition to the two non-equivalent iron sites in the tetragonal structure, an additional magnetic spectral component with a reduced hyperfine field was detected for both substituted alloys, indicating the presence of Re or W defect positions. 
The relative fraction of this substitution-related sextet was shown to be significantly higher for the (Fe$_{0.675}$Co$_{0.275}$Re$_{0.05}$)$_2$B alloy than for the (Fe$_{0.675}$Co$_{0.275}$W$_{0.05}$)$_2$B one. 
Radio frequency Mössbauer studies showed that Re atoms stabilized the amorphous phase, while the alloys (Fe$_{0.7}$Co$_{0.3}$)$_2$B and (Fe$_{0.675}$Co$_{0.275}$W$_{0.05}$)$_2$B began to crystallize when exposed to an rf field of about 12~Oe. 
The higher stability of the Re-substituted amorphous alloy observed in rf-Mössbauer studies is consistent with the results of our thermomagnetic and differential scanning calorimetry measurements. 
As a result of undertaken research novel compositions with high potential due to the high values of magnetization saturation and Curie temperature were characterized. 
In spite of high values of the magnetic anisotropy energy predicted theoretically, the coercivity is limited to about 300~Oe, being still the weakest characteristic of considered materials. 
Nevertheless, coercivity as an extrinsic parameter, still can be improved by e.g. microstructural optimization.

\section*{Acknowledgments}

AM work was supported by the National Science Centre, Poland, within the project No. 2016/23/N/ST3/03820.
MW and WM acknowledge the financial support of the National Science Centre Poland under the decision DEC-2018/30/E/ST3/00267.
M. Kołodziej was financially supported by the project “Środowiskowe interdyscyplinarne studia doktoranckie w zakresie nanotechnologii” (“Environmental interdisciplinary doctoral studies in nanotechnology”) No. POWR.03.02.00-00-I032/16 under the European Social Fund – Operational Programme Knowledge Education Development, Axis III Higher Education for Economy and Development, Action 3.2 PhD Programme.
Part of the computations was performed on the resources provided by the Pozna{\'n} Supercomputing and Networking Center (PSNC).
We thank Justyna Rychły and Justyn Snarski-Adamski for help with language editing and discussion.

\end{sloppypar}

\bibliography{bibliography}

\end{document}